\documentclass[superscriptaddress,twocolumn,aps,prx,10pt]{revtex4-1}

\usepackage{times}
\usepackage{amsmath}
\usepackage{amsfonts}
\usepackage{amssymb}
\usepackage{bm}
\usepackage{graphicx}
\usepackage{color}
\usepackage{float}

\begin{document}
\title{Mode-Coupling Theory of the Glass Transition: A Primer}

\author{Liesbeth M.~C.~Janssen}
\email[Electronic mail: ]{L.M.C.Janssen@tue.nl}
\affiliation{Theory of Polymers and Soft Matter, Department of Applied Physics, Eindhoven University of Technology,
             P.O. Box 513, 5600MB Eindhoven, The Netherlands}

\date{\today}

\begin{abstract}
Understanding the physics of glass formation remains one of the major unsolved
challenges of condensed matter science. As a material solidifies into a glass,
it exhibits a spectacular slowdown of the dynamics upon cooling or compression,
but at the same time undergoes only minute structural changes. Among the
numerous theories put forward to rationalize this complex behavior,
Mode-Coupling Theory (MCT) stands out as the only framework that provides a
fully first-principles-based description of glass phenomenology. This review
outlines the key physical ingredients of MCT, its predictions, successes, and
failures, as well as recent improvements of the theory. We also discuss the
extension and application of MCT to the emerging field of non-equilibrium
active soft matter. 
\end{abstract}

\maketitle

\section{Introduction to the physics of glass formation}
\label{sec:intro}
Glasses are solid materials that lack any long-range structural order,
representing a state of matter that lies somewhere in between a crystalline
solid and a disordered liquid. The most common pathway towards a glassy state
is by rapidly cooling a liquid to below its melting point--thus entering the
so-called supercooled regime--, until the liquid's viscosity $\eta$ simply
becomes so large that it stops flowing on any practical time scale \cite{Debenedetti2001,Berthier2011a,Biroli2013}. The
operational definition of the glass transition temperature $T_g$ is the point where the viscosity
exceeds a value of $10^{12}$ Pa.s or the structural relaxation time $\tau$ exceeds
100 seconds, but most glasses in our everyday lives have a viscosity that is
still orders of magnitude higher \cite{Zanotto1998}. Aside from common
applications such as window panes and household items, amorphous solids can be
found in, e.g., phase-change memory devices, pharmaceutical compounds, optical
fibers, and wearable electronics, and there is compelling evidence that even
living cells employ glass-like behavior to regulate intra- and intercellular
processes \cite{Zhou2009,Angelini2011,Schotz2013a,Sadati2014,Parry2014,Bi2015a,Bi2016}. 
Curiously, most of the water in the universe is also believed to
exist in the glassy state \cite{Debenedetti2003}.

\begin{figure}
        \begin{center}
    \includegraphics[width=0.47\textwidth]{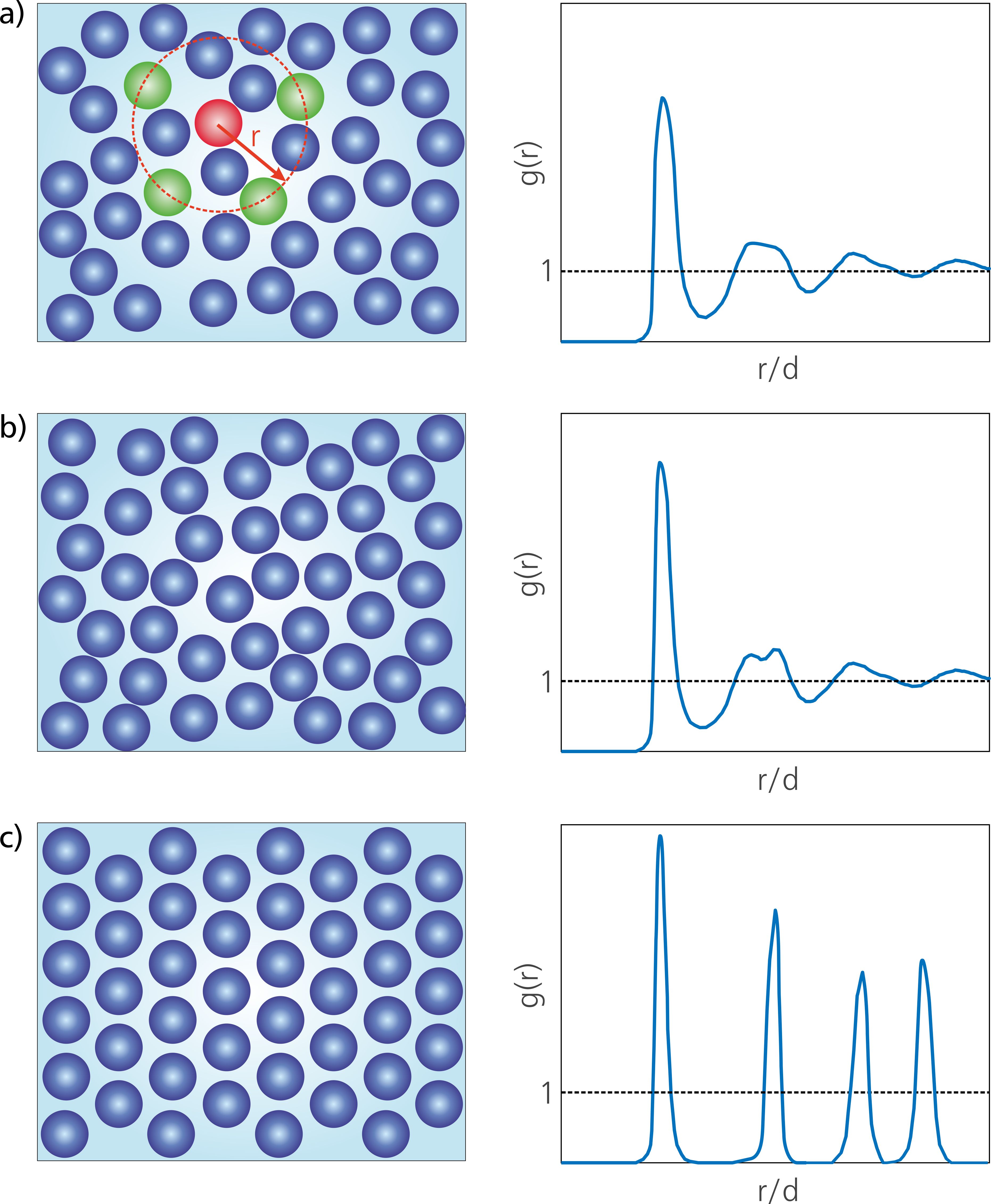}
  \end{center}
  \caption{
  \label{fig:gr}
  Schematic picture of the structure of (a) a normal liquid, (b) an amorphous solid, i.e.\ a glass,
  and (c) a crystalline solid. The first panel highlights four (green) particles which are
  separated by a distance $r$ from a (red) reference particle.
  The right panels illustrate the corresponding
  radial distribution functions $g(r)$, which describe the probability of finding a 
  particle a distance $r$ away from any reference particle, relative to the ideal-gas case.
  The first peak in $g(r)$ represents the first solvation shell at $r\approx 1d$, where $d$ is 
  the particle diameter. The dashed black lines indicate the ideal-gas result $g(r)=1$.
  }
\end{figure}

Given the vast abundance and importance of glasses, it may come as a surprise
that we still understand very little about them. In fact, after decades of
intense research, there is still no consensus on which physical mechanisms
underlie the process of glass formation. Unraveling the nature of the glassy
state ranks among the ''most compelling puzzles and questions facing scientists
today" \cite{Science2005}, and Nobel laureate Philip Anderson even called it ''the deepest and most
interesting unsolved problem in solid-state theory" \cite{Anderson1995}. What makes the glass
transition so notoriously difficult to understand? At the heart of the problem
lies the fact that a vitrifying material exhibits a spectacular growth of
viscosity (or relaxation time) upon cooling or compression, but at the same
time undergoes only minute structural changes. Thus, at the molecular level,
the structure of a glass is almost indistinguishable from that of a normal
liquid (as probed by, e.g., the radial distribution function or the static
structure factor), yet their viscosities differ by at least fifteen (!) orders of
magnitude. This is unlike any conventional thermodynamic phase transition, such
as the liquid-to-crystal transition, which is marked by the appearance of
long-range, periodic structural order (Fig.\ \ref{fig:gr}). Nonetheless, it is not unimaginable that
some kind of 'amorphous order' emerges during vitrification, albeit in a far
less obvious way than in the crystallization example. A popular hypothesis is
that the subtle microstructural changes observed in supercooled liquids might
somehow contain a 'hidden' growing (and possibly diverging) length scale that
accompanies the transition from liquid to amorphous solid, and indeed a large
ongoing effort is devoted to identifying such a length scale \cite{Royall2015,Albert2016}.

\begin{figure}
        \begin{center}
    \includegraphics[width=0.37\textwidth]{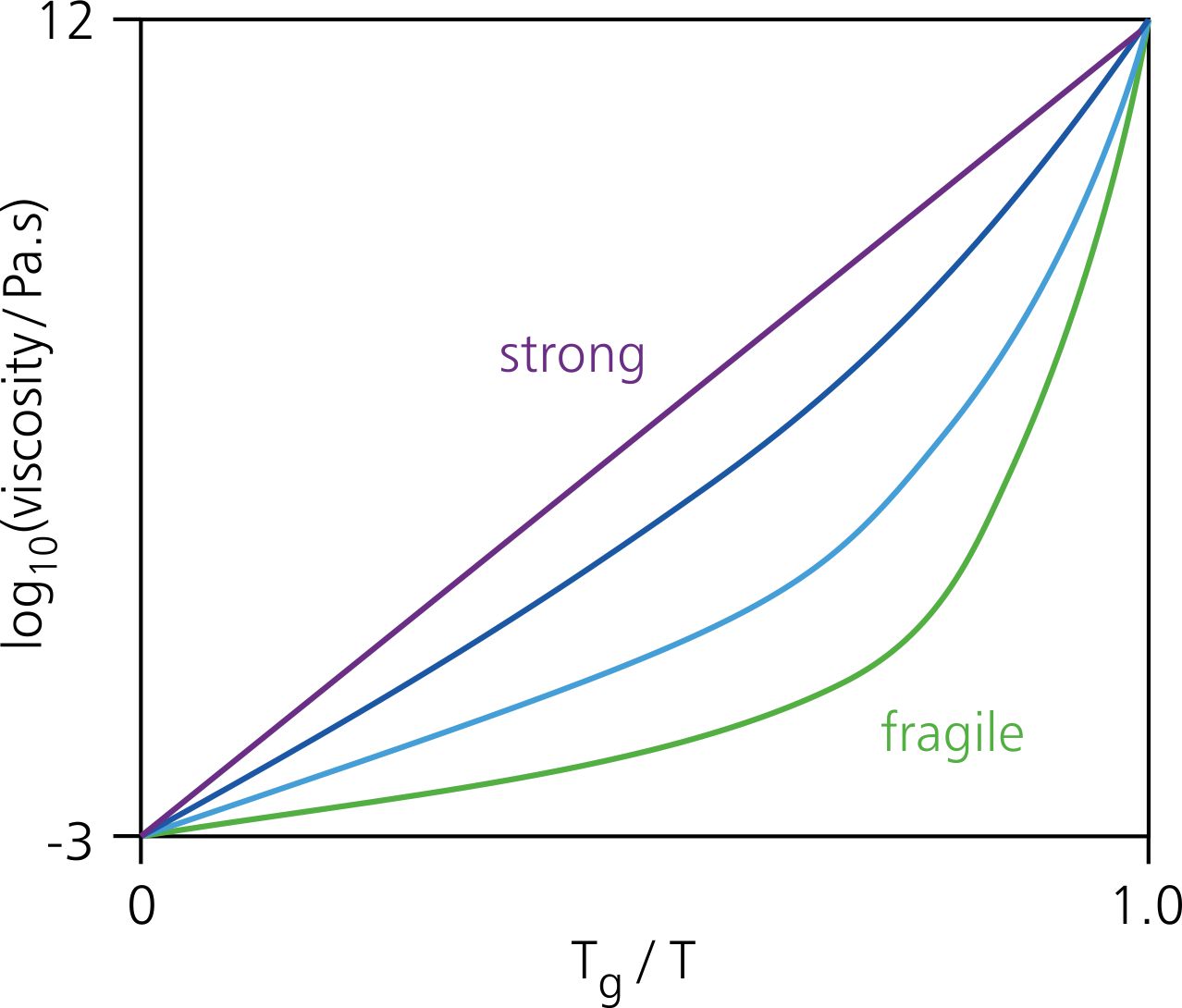}
  \end{center}
  \caption{
  \label{fig:frag}
  Typical fragility plot, showing the logarithm of the viscosity  
  as a function of inverse temperature $1/T$, normalized with respect to the glass transition temperature $T_g$.
  A viscosity of $10^{-3}$ Pa.s corresponds to a normal liquid, while a value of $10^{12}$ Pa.s defines a glassy solid.
  So-called strong glass formers such as silica exhibit an Arrhenius-type growth of the viscosity upon cooling,
  while fragile glass formers such as $o$-terphenyl show a much steeper temperature dependence close to $T_g$.
  Many materials, including colloidal hard spheres and confluent cells, fall in between these two extremes.
  }
\end{figure}

Another major unresolved piece of the glass puzzle is that not all materials
vitrify in the same manner. More specifically, the viscosity growth as a
function of inverse temperature can differ significantly from one material to
another. These differences are captured in an empirical property called
'fragility' \cite{Angell1995,Debenedetti2001}, which characterizes the slope of the viscosity with temperature as
a material approaches the glass transition (Fig.\ \ref{fig:frag}). Materials such as silica fall in
the class of 'strong' glass formers, exhibiting an Arrhenius-type (exponential)
viscosity growth upon cooling, while 'fragile' materials have a viscosity that
increases faster than an Arrhenius law. It is widely believed that a thorough
understanding of the mechanisms underlying fragility will be key to achieving a
universal description of the glass transition, but no theory to date has been
able to predict a material's degree of fragility from the sole knowledge of its
microscopic structure \cite{Tarjus2011}.

While the viscosity already gives an important clue about the complex behavior
of glass-forming materials, the most detailed information is contained in the
microscopic relaxation dynamics, and this will also be the focus of the
remainder of this review. A common probe of such dynamics is the time-dependent
density-density correlation function or so-called intermediate scattering
function, $F(k,t)$, which probes correlations in particle density fluctuations
over a certain wavenumber $k$ and over a time interval $t$ \cite{Hansen2013}. Simply put,
$F(k,t)$ measures to what extent the instantaneous molecular configuration of a
material will resemble the new configuration a time $t$ later; the wavenumber
$k$ designates the inverse length scale over which this resemblance is
measured. By choosing $k$ as approximately one inverse particle diameter,
$F(k,t)$ will thus probe the relaxation dynamics at the molecular level, while
the limit $k\rightarrow 0$ describes the macroscopic dynamics. We note that the
characteristic relaxation time $\tau$ associated with $F(k,t)$ is also a
measure for the viscosity (with the shear modulus as the proportionality
factor \cite{Zanotto1998}), and hence $F(k,t)$ also provides a means to quantify e.g.\ the
fragility.

\begin{figure}
        \begin{center}
    \includegraphics[width=0.48\textwidth]{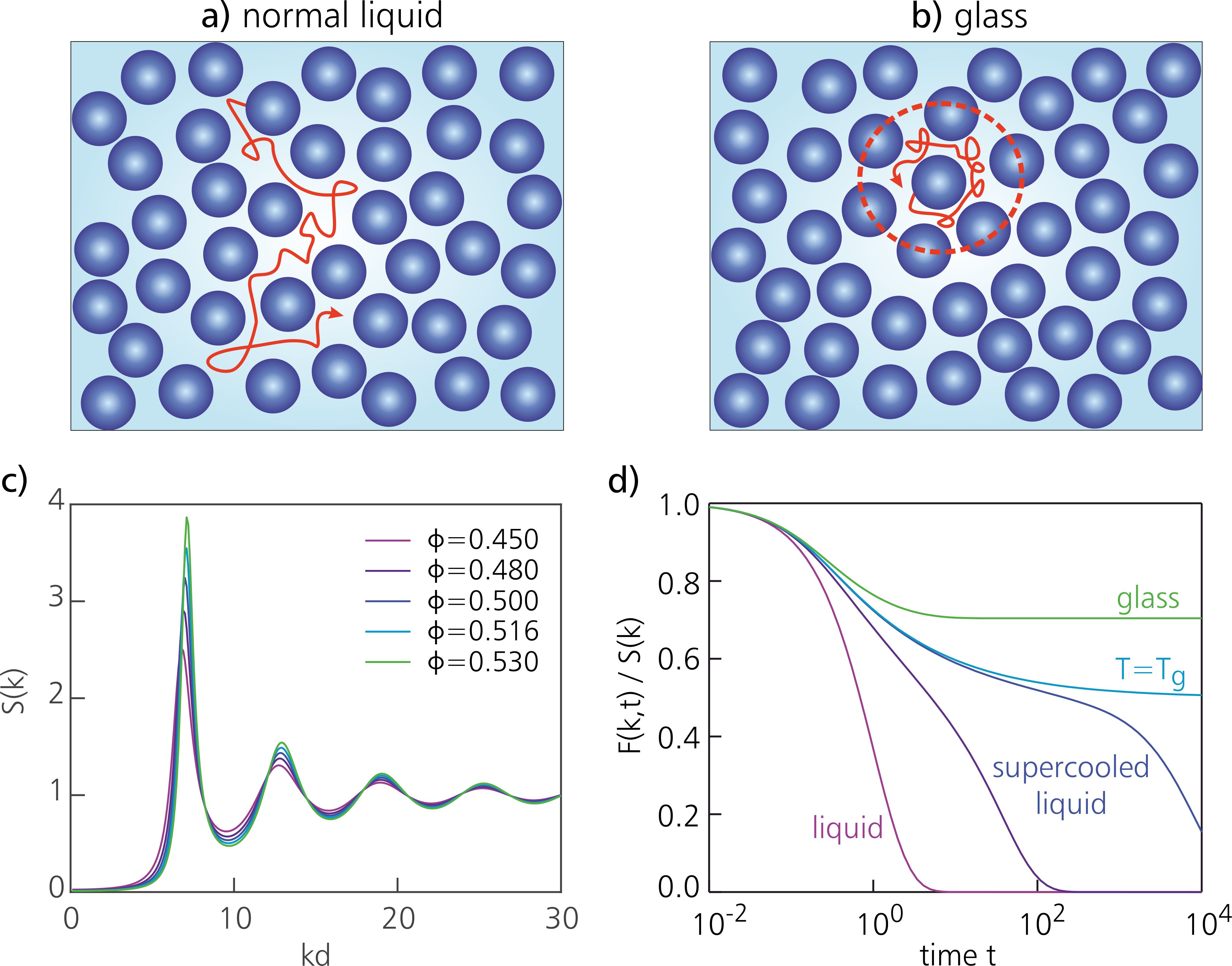}
  \end{center}
  \caption{
  \label{fig:dynamics}
  Schematic picture of the structure and dynamics in a normal liquid, supercooled liquid,
  and glass. Panels (a) and (b) depict a typical trajectory of a particle in the normal liquid phase
  and glassy phase, respectively. In the glassy state, 
  particles become trapped in a cage formed by their neighbors. The dashed red line indicates 
  the typical size of a cage, with a radius of approximately one particle diameter $d$. 
  Panel (c) shows the static structure factors $S(k)$ for a glass-forming system
  of hard spheres for several packing fractions $\phi$, calculated using the Percus-Yevick approximation.
  The main peak position of $S(k)$ 
  corresponds to a wavenumber of approximately one inverse particle diameter, $k \approx 2\pi/d$.
  Within MCT, the glass transition for this system takes place at $\phi_c=0.516$ \cite{Bengtzelius1984}.
  Panel (d) shows typical intermediate scattering functions $F(k,t)$ as a function of time 
  for $k \approx 2\pi/d$. As the temperature is decreased or the packing fraction is increased,
  the system becomes more glassy and $F(k,t)$ decays more slowly. 
  }
\end{figure}

The behavior of $F(k,t)$ upon cooling thus reveals how the microscopic
relaxation dynamics changes during the vitrification process \cite{Reichman2005,Kob2002} (Fig.\ \ref{fig:dynamics}). 
In a normal
high-temperature liquid, $F(k,t)$ will decay to zero in a rapid and simple
exponential fashion, since the particles can move around easily and therefore
quickly lose track of their initial positions.
At temperatures in the supercooled regime, however, $F(k,t)$ shows a more
complex multi-step relaxation pattern (also see Fig.\ \ref{fig:Fkt_MCT}): at intermediate times (the so-called
$\beta$-relaxation regime), a plateau develops during which $F(k,t)$ remains
constant, indicating the transient freezing of particles; only at sufficiently
long times will the correlation function fully decay to zero. Notably, this
final decorrelation process (so-called $\alpha$-relaxation) is not a simple
exponential decay, as in a normal liquid, but rather a more slowly decaying,
'stretched' exponential behavior of the form $\exp(-t/\tau)^{\beta}$, with
$0<\beta<1$. As the temperature decreases toward the glass transition
temperature, the plateau in $F(k,t)$ will extend to increasingly long times,
until it finally exceeds the entire time window of observation. Thus, at the
glass transition, $F(k,t)$ fails to decorrelate on any practical time
scale--implying that particles always stay reasonably close to their initial
positions--, marking the onset of solidity. The final value of the intermediate
scattering function, $f(k)=\lim_{t\rightarrow \infty} F(k,t)$, is known as the
non-ergodicity parameter \cite{Megen1991}, and is often used as the order parameter for the
glass transition: $f(k)=0$ corresponds to the liquid state, and $f(k)>0$
indicates a solid (Fig.\ \ref{fig:dynamics}).

\begin{figure}
        \begin{center}
    \includegraphics[width=0.43\textwidth]{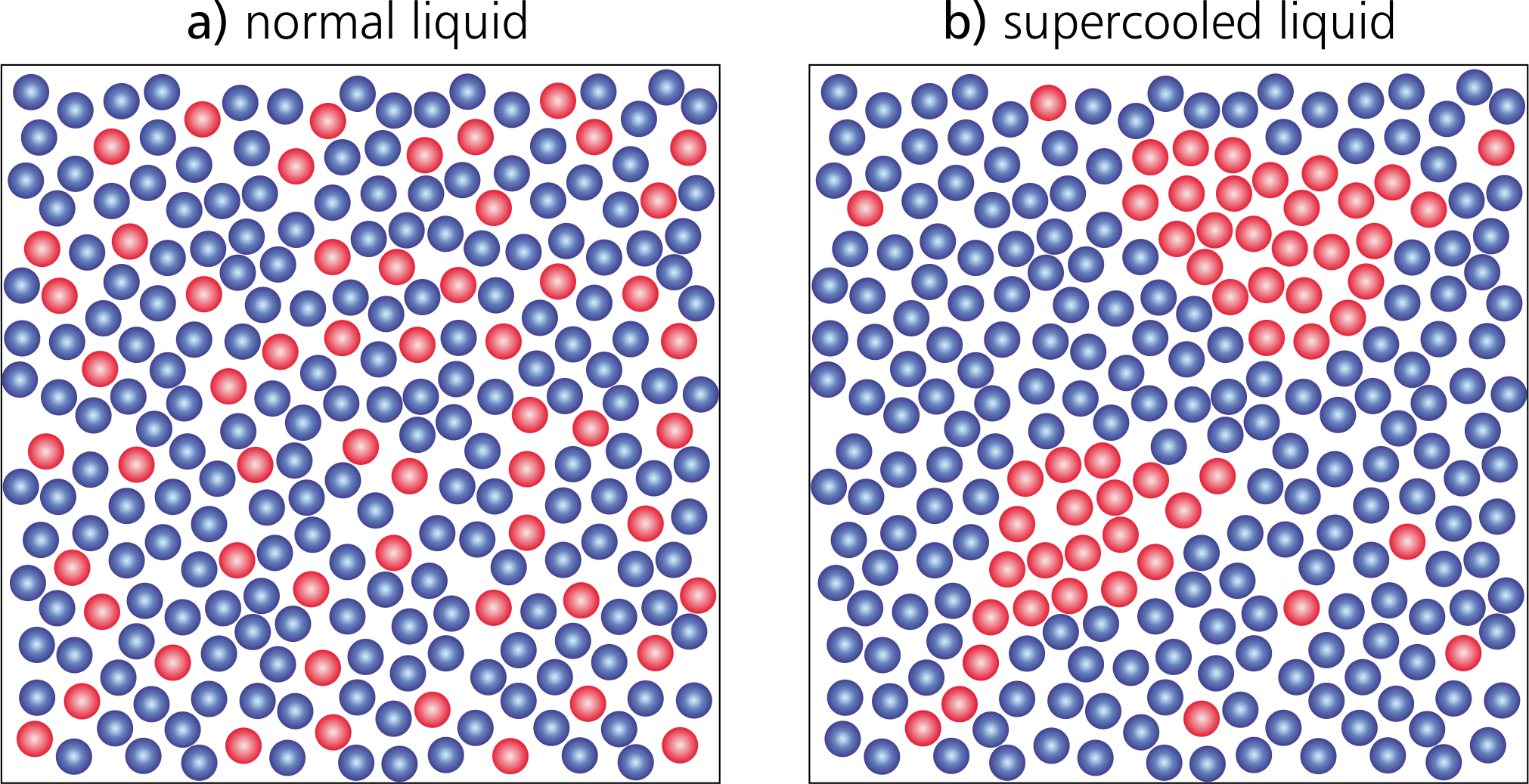}
  \end{center}
  \caption{
  \label{fig:DH}
  Illustration of dynamic heterogeneity in supercooled liquids. The red-colored particles
  represent mobile particles that have moved further than a certain distance $\Delta r$
  during a time interval $\Delta t$, while blue-colored particles represent particles that 
  have moved less than this distance
  in the same time interval. In the normal liquid phase (a), particle motion occurs rather homogeneously
  across the entire sample. Conversely, in a supercooled liquid (b), particle motion occurs heterogeneously
  in clusters of collectively moving particles, and the appearance of such mobile clusters fluctuates both
  in space and in time. The figure is based on Ref.\ \cite{Weeks2000}.
  }
\end{figure}

There are several other aspects in the dynamics of supercooled liquids that
differ markedly from those seen in ordinary liquids, including the emergence of
dynamic heterogeneity \cite{Weeks2000,Kegel2000,Ediger2000,Berthier2011}
and the breakdown of the Stokes-Einstein relation \cite{Tarjus1995,Shi2013}.
Dynamic heterogeneity refers to the fact that structural relaxation does not
take place uniformly throughout the entire material--as in a normal liquid--,
but rather in clusters of collectively rearranging particles, while the rest of
the supercooled liquid remains temporarily frozen (Fig.\ \ref{fig:DH}). The appearance of such
mobile domains will vary both in space and in time, thus giving rise to
non-trivial spatiotemporal fluctuations that become more pronounced as the
glass transition is approached. Dynamic heterogeneity cannot be seen in
$F(k,t)$ itself, but rather in the \textit{fluctuations} of $F(k,t)$ among
different particle trajectories \cite{Biroli2004}. These fluctuations are encoded in the
so-called dynamic susceptibility $\chi_4(t)$, whose peak height is a measure
for the size of the cooperatively rearranging regions. As a material is
supercooled, a growing $\chi_4(t)$ thus indicates a growing dynamic length
scale associated with vitrification, but a true divergence of this length
scale--as expected for typical critical phenomena--has not yet been observed \cite{Albert2016}. A
related puzzling phenomenon concerns the Stokes-Einstein equation, which states
that the viscosity (or relaxation time), diffusion constant $D$, and
temperature of a liquid are related as $D\eta / T = \rm{constant}$. This ratio
holds generally for normal liquids, but in the supercooled regime the viscosity
increase tends to be stronger than the diffusion-constant decrease. This
breakdown of Stokes-Einstein behavior is widely believed to be a manifestation
of dynamic heterogeneity, but the fundamental origins of both phenomena remain
poorly understood.

In this review, we focus on one of several theories that seeks to describe the
above complex phenomenology of glass-forming materials, namely Mode-Coupling
Theory (MCT) \cite{Gotze2008,Das2004}. This theory was first put forward by G\"{o}tze and coworkers in
the 1980s \cite{Bengtzelius1984,Leutheusser1984}, and continues to stand out as the \textit{only} framework of glassy
dynamics that is based entirely on first principles, starting from the exact
microscopic picture of a correlated liquid. We outline the key physical
ingredients and sketch of the MCT derivation, its predictions, successes, and
failures, as well as recent improvements and extensions of the theory. Part of
this work is based on the review by Reichman and Charbonneau \cite{Reichman2005} 
and by Szamel \cite{Szamel2013}; for a detailed
discussion of the original theory, including an extensive treatment of the
involved mathematics, we refer to the seminal work of G\"{o}tze \cite{Gotze2008}. For an
overview of the many other existing theories of glass formation, see e.g.\
Refs.\ \cite{Tarjus2011,Berthier2011a,Langer2014}.

\section{Derivation of the MCT equations}
\label{sec:deriv}

\subsection{Preliminaries}

As already noted in the introduction, MCT provides the only first-principles
route towards the description of glassy behavior, making it a unique theory
that does not rely on any phenomenological assumptions. Explicitly, MCT aims to
predict the full microscopic relaxation dynamics of a glass-forming material--as
a function of time, wavenumber, temperature, and density--, using only knowledge of
static, time-independent properties as input. Aside from constants such as the
system's temperature and density, the main theory input is the average
microscopic structure of the material. The simplest experimental measure of the latter 
is the static structure factor $S(k)$, which can be obtained directly from scattering experiments.
This structure function is related to the radial distribution function $g(r)$ through a Fourier transform \cite{Hansen2013,Zhang2016},
and thus probes--in Fourier space--the likelihood of finding a particle at a certain distance $r \sim 2\pi/k$ away from any
other particle (Fig.\ \ref{fig:dynamics}). Formally $S(k)$ is also
 equivalent to $F(k,t=0)$. It must be noted that MCT also admits more
intricate three-particle correlation functions as additional structural input,
but--with the exception of network-forming fluids \cite{Sciortino2001,Coslovich2013}--the sole knowledge of $S(k)$
generally suffices. Importantly, it is through these structural metrics that
MCT knows about the chemical composition of the material under study. That is,
the theory is able to distinguish between, say, a glass-forming fluid of silica
or Lennard-Jones particles only through their differences in (wavevector-dependent) structure.

\begin{figure}
        \begin{center}
    \includegraphics[width=0.42\textwidth]{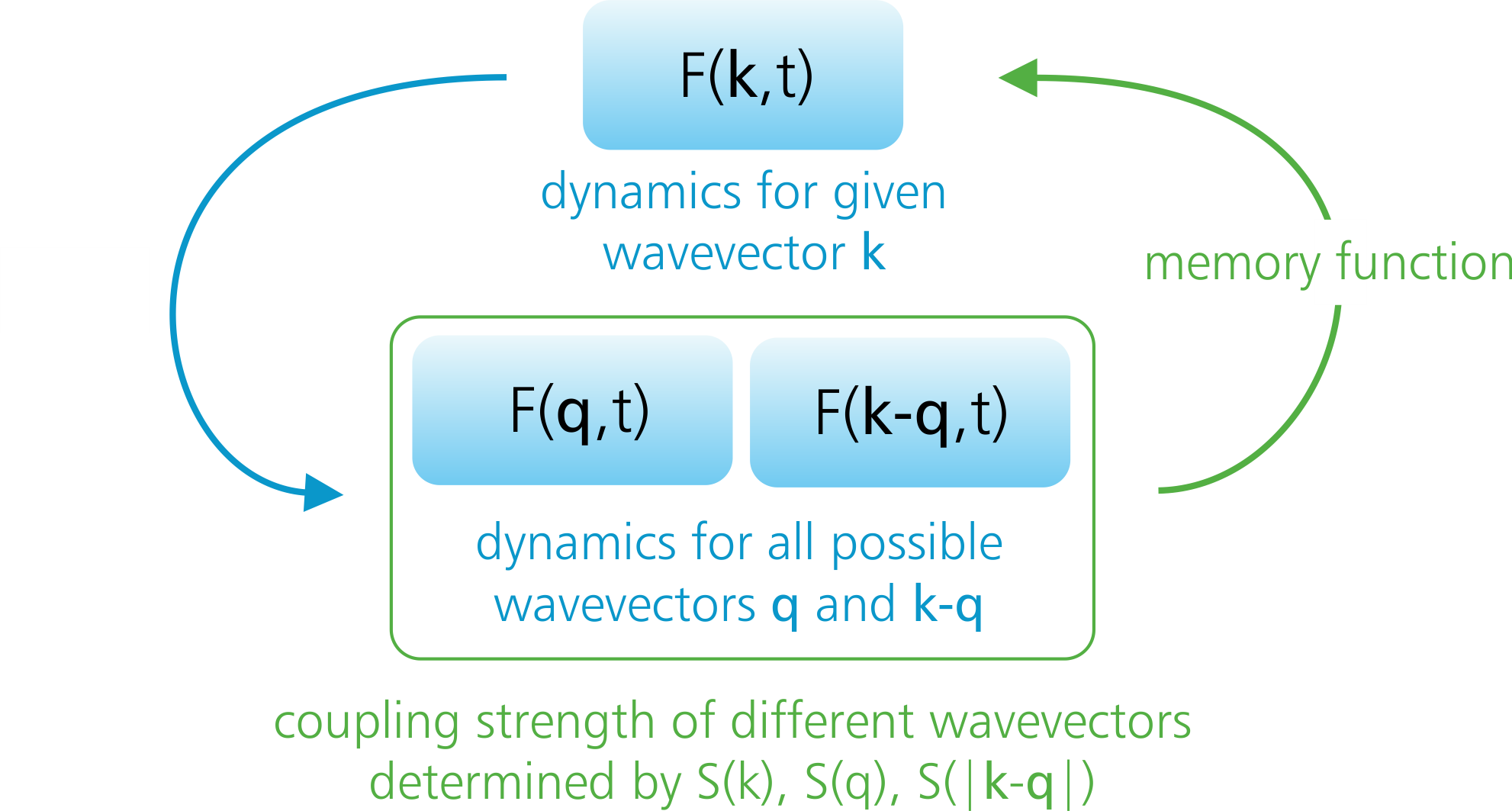}
  \end{center}
  \caption{
  \label{fig:MCT}
  Sketch of the MCT equations. The theory seeks to predict the full dynamics of the intermediate
  scattering function $F(k,t)$ for all possible wavevectors $\mathbf{k}$ and all times $t$. 
  The exact $F(k,t)$ dynamics is governed by a memory function that, within standard MCT, is 
  approximated as a product of two intermediate scattering functions that probe density correlations
  at \textit{different} wavevectors $\mathbf{q}$ and $\mathbf{k-q}$. The couplings among the 
  different wavevectors $\{\mathbf{k}, \mathbf{q}, \mathbf{k-q} \}$
  are determined by the so-called vertices, which depend explicitly on the static structure factor. Hence,
  the static structure factor must be given as input to the theory, and dynamical information is given 
  as output.
  }
\end{figure}

In the standard formulation of MCT, the theory seeks to predict the full
dynamics of the intermediate scattering function $F(k,t)$ of a given material,
starting with the \textit{exact} equation of motion for $F(k,t)$. Below we
sketch the derivation of this equation, followed by a discussion of the various
MCT approximations made to solve it. Briefly, the derivation will amount to an
exact integro-differential equation for $F(k,t)$ [Eq.\ (\ref{eq:memeq})] that
is governed by an even more complicated time-dependent correlation ('memory')
function. MCT makes the \textit{ad hoc} assumption that the latter memory
function can be approximated as a product of $F(k,t)$ functions, thus yielding
a closed, self-consistent equation (see Fig.\ \ref{fig:MCT}).  As described in
Sec.\ \ref{sec:MCT}, the final MCT equation [Eq.\ (\ref{eq:MCT})] is
reminiscent of a damped harmonic oscillator, but with a \textit{time-dependent}
damping term that ultimately produces the dramatic dynamic slowdown in
supercooled liquids.

Let us first define our variables of interest,
namely the collective density modes,
\begin{eqnarray}
\label{eq:rho}
\rho(\mathbf{r},t) &=& \sum_j^N \delta(\mathbf{r} - \mathbf{r}_j(t)), \\ \nonumber
\rho(\mathbf{k},t) &=& \int d\mathbf{r} e^{i\mathbf{k}\mathbf{r}} \rho(\mathbf{r},t) \\ \nonumber
&=& \sum_j^N e^{i\mathbf{k}\mathbf{r}_j(t)}, 
\end{eqnarray}
where $N$ denotes the total number of particles and $\mathbf{r}_j(t)$ is the
position of particle $j$ at time $t$. The real-space density
$\rho(\mathbf{r},t)$ thus simply measures where all particles are located at a
given point in time, and $\rho(\mathbf{k},t)$ is the corresponding Fourier
transform for wavevector $\mathbf{k}$. The intermediate scattering function
$F(k,t)$ probes the time-dependent correlations between these collective
density modes,
\begin{equation}
\label{eq:Fkt}
F(k,t) = \frac{1}{N} \langle \rho(-\mathbf{k},0) \rho(\mathbf{k},t) \rangle,
\end{equation}
where the brackets denote a canonical ensemble average. 
At time $t=0$, this correlation function reduces to the static structure factor,
\begin{equation}
S(k) = \frac{1}{N} \langle \rho(-\mathbf{k},0) \rho(\mathbf{k},0) \rangle \equiv F(k,0),
\end{equation}
which thus contains information on the static density distribution of the
material, i.e., the average microscopic structure. Note that in an isotropic
material, such as a powder or a 'simple' fluid, both $S(k)$ and $F(k,t)$ depend
only on the magnitude of the wavector, $k=|\mathbf{k}|$, but in e.g.\ the
presence of an external field the full wavevector dependence should be
considered \cite{Biroli2006}.
 
\subsection{Mori-Zwanzig projection formalism}
\label{sec:MZ}

In order to obtain an exact equation of motion for $F(k,t)$, we make use of the
so-called Mori-Zwanzig projection formalism \cite{Zwanzig1961,Mori1965}. The basic idea behind this
formalism is to divide the entire universe into two mutually orthogonal
subspaces: one containing the variables of interest, and one simply containing
'everything else'. The goal is to describe how the dynamics of the relevant
variables evolves over time, in the presence of all other 'non-interesting' variables.
Here we will focus mainly on molecular glass-forming fluids,
in which case the variables of interest are the collective density
modes of Eq.\ (\ref{eq:rho}) and their associated current modes 
\begin{equation}
\label{eq:jk}
j(\mathbf{k},t) \equiv \dot{\rho}(\mathbf{k},t) = i \sum_{l=1}^N (\mathbf{k} \cdot \dot{\mathbf{r}}_l) e^{i\mathbf{k}\mathbf{r}_l(t)},
\end{equation}
where the dots denote time derivatives. Note that in general there is no simple
recipe for deciding which variables are `relevant'; typically we focus on
quasi-conserved or 'slow' variables that show some non-trivial time-dependence
(unlike strictly conserved variables that are constant), but which do not
fluctuate too fast either, so as not to be confused with noise. From Eq.\
(\ref{eq:jk}), it is easy to see that in the limit $k\rightarrow 0$,
corresponding to very large length scales, the current $\dot{\rho}(\mathbf{k},t)$ will
vanish and consequently the macroscopic density is strictly conserved. On
smaller length scales, however, i.e., $k > 0$, the local density will
fluctuate as particles move around, and it is these fluctuations--and their
time-dependent correlations--that we seek to probe in $F(k,t)$ and predict with
MCT.

For convenience we will organize the variables $\rho(\mathbf{k},t)$ and $j(\mathbf{k},t)$ into a
two-component vector $\underline{A}$, which thus spans our subspace of
interest:
\begin{equation}
\label{eq:A}
\underline{A}(t) \equiv \begin{bmatrix} A_1(t) \\ A_2(t) \end{bmatrix} = \begin{bmatrix} \rho(\mathbf{k},t) \\ j(\mathbf{k},t) \end{bmatrix}.
\end{equation}
Importantly, in this notation, time-dependent correlation functions may now be
identified simply as \textit{scalar products} between such vector elements,
e.g., $F(k,t) = (1/N) \langle A_1(0) | A_1(t) \rangle = (1/N) \langle A_1^*(0)
A_1(t) \rangle$, where we have used the standard bra-ket notation with the
asterisk representing the complex conjugate.  We define the full matrix of all
possible scalar products as $\underline{C}(t)$, with matrix elements 
\begin{equation}
C_{\alpha\beta}(t) \equiv \langle A_{\alpha}(0) | A_{\beta}(t) \rangle.
\end{equation}
Note that the first matrix element $C_{11}(t)$ equals $NF(k,t)$, and
$C_{21}(t)=(N/i)(dF(k,t)/dt)$. Furthermore, in analogy to ordinary projections
in vector space, we can now use these scalar products to define a
\textit{projection operator} $\mathcal{P}_A$ as
\begin{equation}
\mathcal{P}_A = \sum_{\alpha,\beta} |A_{\alpha}\rangle [\underline{C}(0)^{-1}]_{\alpha\beta} \langle A_{\beta} | 
\end{equation}
where the sums run over all possible matrix elements. The projection of some
vector $\underline{X}$ onto $\underline{A}$ is then given by
$\mathcal{P}_A\underline{X}$. Such a projection essentially extracts all the
'slow' or 'relevant' character (defined through $\underline{A}$) from an
arbitrary variable $\underline{X}$, leaving the remaining part of
$\underline{X}$ orthogonal to $\underline{A}$. It is easy to
show that $\mathcal{P}^2_A=\mathcal{P}_A$ and
$\mathcal{P}_A\underline{A}=\underline{A}$, i.e., the projection of
$\underline{A}$ onto itself returns $\underline{A}$. This projection formalism,
introduced by Zwanzig and Mori, thus establishes a link between dynamic
variables and standard vector algebra. Without any loss of generality, it will enable us to separate the full
dynamical behavior of our system into two contributions: i) the dynamics
evolving in the 'slow' subspace spanned by $\underline{A}(0)$, and ii) the
dynamics due to all remaining 'fast' variables, obtained simply by projecting
out all the slow $\underline{A}$-character from the full dynamics. 

Let us now look explicitly at the time-dependent dynamics of a glass-forming
supercooled liquid.  For classical fluids that obeys Newton's equation of
motion, the time evolution of $\underline{A}(t)$ can always be formally written
as
\begin{equation} 
\label{eq:EOM_formal}
\underline{A}(t)=e^{i\mathcal{L}t} \underline{A}(0),
\end{equation} 
where $\mathcal{L}$ is the so-called Liouvillian operator.  The definition of
$\mathcal{L}$ can be found in, e.g., Ref.\ \cite{Reichman2005}, but here we
will not be concerned with its explicit form; it suffices to know that this
operator governs the full dynamics of our variables of interest.  Note that for
colloidal glass-forming systems undergoing Brownian rather than Newtonian
motion, a similar equation applies when considering only the density modes in $\underline{A}$ and 
replacing the Liouvillian by the so-called Smoluchowski operator \cite{Nagele1996}.

While Eq.\ (\ref{eq:EOM_formal}) is formally exact, it does not necessarily
yield any new physical insight into the complex time-dependent dynamics of
supercooled liquids.  Instead, we can rewrite this equation through a somewhat
lengthy derivation (involving the insertion of the unit
matrix operator $1=\mathcal{P}_A+1-\mathcal{P}_A$ and separating the
time-evolution operator $\exp{(i\mathcal{L}t)}$ into a 'slow' component and its
orthogonal part) in the following form \cite{Reichman2005}:
\begin{equation}
\label{eq:GLE}
\frac{d\underline{A}(t)}{dt} = i\underline{\Omega}\cdot\underline(A)(t) - \int_0^t ds \underline{K}(s) \cdot \underline{A}(t-s) + \underline{f}(t).
\end{equation}
For the matrix of correlation functions $\underline{C}(t)$ we similarly find
\begin{equation}
\label{eq:memeq}
\frac{d\underline{C}(t)}{dt} = i\underline{\Omega}\cdot\underline{C}(t) - \int_0^t ds \underline{K}(s) \cdot \underline{C}(t-s).
\end{equation}
Here, $\underline{\Omega}$ is the so-called frequency matrix (the name will
become apparent later on), which captures the part of the time derivative of
$\underline{A}$ that remains in the slow subspace as time evolves,
$\underline{K}(s)$ is a time-dependent \textit{memory function}, and
$\underline{f}(t)$ is the 'fast' \textit{fluctuating force}, which is defined
as
$\underline{f}(t)=e^{i(1-\mathcal{P}_A)\mathcal{L}t}i(1-\mathcal{P}_A)\mathcal{L}\underline{A}(0)$.
That is, $\underline{f}(t)$ is obtained by first removing all the 'slow'
character from the time derivative of $\underline{A}$ using the complementary
projection operator $(1-\mathcal{P}_A)$, and is subsequently propagated in time
in the 'fast' subspace orthogonal to $\underline{A}$. The memory function
$\underline{K}(t)$ is given by the time-autocorrelation function of this
fluctuating force; physically, $\underline{K}(t)$ represents a dissipative term
that ultimately breaks the conservation of $\underline{A}$. In other words,
$\underline{K}(t)$ and $\underline{f}(t)$ embody how our slow variable
$\underline{A}$--which at time $t=0$ lives strictly in the slow subspace--will
gradually evolve under the influence of the rest of the universe, e.g.\ in the
presence of 'fast' variables such as thermal noise. Note that in arriving at
Eq.\ (\ref{eq:memeq}), we have used that $\langle \underline{A}(0) |
\underline{f}(t) \rangle=0$ by construction.  Importantly, Eqs.\ (\ref{eq:GLE})
and (\ref{eq:memeq}), which are known as the generalized Langevin equation and
memory equation, respectively, are both \textit{exact}.

\subsection{Mode-Coupling Theory approximations}
\label{sec:MCT}

By Eq.\ (\ref{eq:memeq}), the difficulty of predicting the full time-dependent
dynamics of $F(k,t)$ is now deferred to the the question of how the memory
function $\underline{K}(t)$ evolves with time. In general, there is no rigorous solution
for this equation, and hence approximations must be made. The main idea behind
MCT is to approximate $\underline{K}(t)$ in 'the simplest non-trivial way' using a two-step
approach:

\textbf{1. Approximate the memory function as a four-point density correlation function.}
First, using the density modes as the main physical variables of interest,
the fluctuating force $\underline{f}(t)$ is projected onto a new basis of
\textit{products of two density modes},
$\rho(\mathbf{k}_1,t)\rho(\mathbf{k}_2,t)$, where $\mathbf{k}_1$ and
$\mathbf{k}_2$ run over all possible wavevectors relevant to our system.
Physically, this projection is motivated by the fact that for particles
interacting through an arbitrary pair potential, such products of densities
emerge naturally in the expression for the fluctuating force \cite{Reichman2005}. This may seem
rather counterintuitive at first, since the fluctuating force is a fast
variable while density modes are slow by definition, but it can be shown by
Fourier transformation that, for an $n$-body interaction potential,
$\underline{f}(t)$ always contains products of $n$ density modes \cite{Schilling2005}.  In the
standard MCT formulation, it is assumed that the pair densities dominate the
fluctuating force entirely, but higher-order generalizations with projections
onto an $n$-density-mode basis have also been considered
\cite{Schofield1992,Zon2001}.  Mathematically, the projection onto pair densities also
corresponds to the first non-vanishing component in density space, i.e., 'the simplest non-vanishing term', since a
projection onto a single density mode would always give zero by construction \cite{Gotze2008}.
Overall, this approximation brings the memory function $\underline{K}(t)$,
which is the time-correlation function of $\underline{f}(t)$, into the form of
a \textit{four-point density correlation function}:
\begin{eqnarray}
\underline{K}(t) &\sim& \sum_{\mathbf{k}_1,\mathbf{k}_2,\mathbf{k}_3,\mathbf{k}_4} \\ \nonumber
& & \langle \rho^*(\mathbf{k}_1,0) \rho^*(\mathbf{k}_2,0) e^{i[1-\mathcal{P}_A]\mathcal{L}t} \rho(\mathbf{k}_3,0) \rho(\mathbf{k}_4,0) \rangle,
\end{eqnarray}
with the time-propagation operator $\exp{[i(1-\mathcal{P}_A)\mathcal{L}t]}$
acting in the fast subspace.

\textbf{2. Factorize four-point correlation functions into two-point correlation functions.}
Second, the (unknown) four-point correlation functions in $\underline{K}(t)$
are further simplified by factorizing them into a product of two two-point
correlation functions $\langle
\rho^*(\mathbf{k}_1,0)\rho(\mathbf{k}_1,t)\rangle$ and $\langle
\rho^*(\mathbf{k}_2,0)\rho(\mathbf{k}_2,t)\rangle$. At the same time, the
operator $\exp{[i[1-\mathcal{P}_A]\mathcal{L}t]}$ is replaced by the normal
operator $\exp{[i\mathcal{L}t]}$, since the single density modes
$\rho(\mathbf{k}_1,t)$ and $\rho(\mathbf{k}_2,t)$, which start out in the slow
subspace, would otherwise give a zero contribution. It is important to note
that this factorization is an \textit{ad hoc} approximation that is not
necessarily motivated by any physical insight; rather, it merely serves to
produce a 'simple' memory function that is not trivially zero.  Nonetheless, it
can be shown that the factorization is exact for so-called Gaussian variables \cite{Zaccarelli2001},
but density modes in general do not behave as such. 

After the second approximation is made, we may then realize that the factorized
two-point density correlation functions $\langle
\rho^*(\mathbf{k}_i,0)\rho(\mathbf{k}_i,t)\rangle$ are, in fact, equal to
$NF(k_i,t)$ by virtue of Eq.\ (\ref{eq:Fkt}). Thus, our full equation of motion
for the intermediate scattering function $F(k,t)$ is now governed by a memory
function containing precisely the same function, but for many different
wavenumbers. After explicitly working out all the expressions for the frequency
matrix $\underline{\Omega}$ and the (approximate) memory function
$\underline{K}(t)$, and concentrating on the lower left corner of the
correlation matrix $C_{22}(t)$ in Eq.\ (\ref{eq:memeq}), we finally arrive at
the full MCT equation \cite{Reichman2005}:
\begin{equation}
\label{eq:MCT}
\frac{d^2F(k,t)}{dt^2} + \frac{k_BTk^2}{mS(k)}F(k,t) + \int_0^t ds K_{MCT}(s) \frac{dF(k,t-s)}{dt} = 0,
\end{equation}
with the memory function given by
\begin{equation}
\label{eq:MCTmem}
K_{MCT}(t)= \frac{\rho k_BT}{16\pi^3m} \int d\mathbf{q} |V_{\mathbf{q},\mathbf{k-q}}|^2 F(q,t) F(|\mathbf{k-q}|,t).
\end{equation}
Here, $k_B$ is the Boltzmann constant, $m$ is the particle mass, $\rho$ is the
bulk density, and the factors 
\begin{equation}
V_{\mathbf{q},\mathbf{k-q}}=k^{-2}[(\mathbf{k}\cdot\mathbf{q})c(q)+\mathbf{k}\cdot(\mathbf{k-q})c(|\mathbf{k-q}|)]
\end{equation}
are referred to as vertices, with $c(k)=\rho^{-1}[1-1/S(k)]$ denoting the
direct correlation function \cite{Hansen2013}. These vertices represent the strength of the
coupling between different density modes at wavevectors $\mathbf{q}$ and
$\mathbf{k-q}$.  In arriving at this equation, we have also assumed that $S(k)$
contains all the relevant microscopic structural information (using the
so-called convolution approximation \cite{Jackson1962,Kob2002, Reichman2005}), but in general the vertices may also
contain higher-order, triplet-density correlations \cite{Sciortino2001,Berthier2009a}. 
 Equation (\ref{eq:MCT}) is
a closed, \textit{self-consistent} equation, and is subject to the boundary
conditions $F(k,0)/S(k)=0$ and $\dot{F}(k,0)=0$.

Let us briefly compare this MCT result with the equation of motion for a
one-dimensional damped harmonic oscillator:
$\ddot{x}+\omega^2x+2\zeta\omega\dot{x}=0$, where $\omega$ is the frequency of
the undamped oscillator and $\zeta$ is the damping coefficient.  It can be seen
that the MCT equation is rather similar, with
$\underline{\Omega}_{22}=k_BTk^2/[mS(k)]$ playing the role of $\omega^2$.
Hence, the $\underline{\Omega}$ matrix is referred to as the frequency matrix.
The damping coefficient, on the other hand, appears in the MCT equation in the
form of the memory function $K_{MCT}(t)$ (note the first derivative of $F(k,t)$
in the integrand). Consequently, we may interpret the memory function as a
generalized, time-dependent damping, which will ultimately cause the dynamical
slowdown in $F(k,t)$ \cite{Kob2002}.

While analytic solutions of the MCT equation generally do not exist, it is
always possible to solve the equation numerically, namely by iteratively making
an ansatz for $F(k,t)$ for all $k$, subsequently constructing the memory
function, and updating $F(k,t)$ until convergence is reached.  We also note
that for systems undergoing Brownian instead of Newtonian dynamics, in which
case the Liouvillian should be replaced by the Smoluchowski operator, MCT
yields an \textit{identical} equation (with $k_BT/m$ being replaced by the
diffusion constant $D$) \cite{Szamel1991}; however, the origin of this similarity is subtle and
rather non-trivial \cite{Voigtmann}.  Moreover, it has also been
shown that this equation applies reasonably well to glass-forming polymers
\cite{Chong2007}, suggesting that MCT captures at least some degree of
universal dynamical behavior. Finally, we note that MCT-based equations have also
been formulated for, e.g., the stress correlation function, the dynamics under
shear deformation, and microrheology studies, but these will not further be discussed
in this review.

\section{Mode-Coupling Theory predictions}
 
The microscopic MCT equation, Eq.\ (\ref{eq:MCT}), can be solved for any
glass-forming material at a given bulk density $\rho$ and temperature $T$ once
the corresponding static structure factor $S(k)$ is known. Thus, MCT predicts
the full microscopic dynamics given only time-independent information as input.
In order to describe the entire vitrification process from liquid to glass, one
typically measures $S(k)$ for a series of temperatures or densities, and
performs a separate MCT calculation for every relevant temperature and density.
In this section, we summarize the main successes and failures of such MCT
predictions.

\subsection{Successes}
\label{sec:successes}
Despite the various approximations made in MCT, the theory gives a remarkable
set of accuracte predictions. Firstly, MCT is indeed capable of predicting a
glass transition, which is non-trivial considering that the static structure
factor $S(k)$--the main theory input--changes only very weakly upon
vitrification (Fig.\ \ref{fig:dynamics}c). As mentioned earlier, the relaxation time of the predicted $F(k,t)$ is used as an
indicator for the glassiness: at the glass transition, the relaxation time
diverges and $F(k,t)$ fails to decay to zero on any time scale. The corresponding
non-ergodicity parameter $f(k)$ is also often in good quantitative agreement with 
the results of computer simulations and experiments (see e.g.\ Refs.\ \cite{Kob2002a,Weysser2010}).

Mathematically,
MCT's ability to predict a glass follows from the non-linearity of the equation
(by virtue of the product of two $F(k,t)$ functions in the memory function),
which renders the theory very sensitive to any small change in structural
input. This non-linearity leads to a feedback mechanism that ultimately drives
the dramatic dynamical slowdown: upon cooling, $S(k)$ will become slightly
larger at certain wavevectors, causing the vertices to increase as well.
Consequently, the memory function will become larger and produce a stronger
damping for $F(k,t)$. The resulting slower intermediate scattering function
will further strengthen the memory function, slowing down the dynamics even
more. This non-linear feedback effect explains at least qualititatively why the
relaxation dynamics can change so dramatically upon only small changes in the
structure and temperature \cite{Kob2002}. 

\begin{figure}
        \begin{center}
    \includegraphics[width=0.48\textwidth]{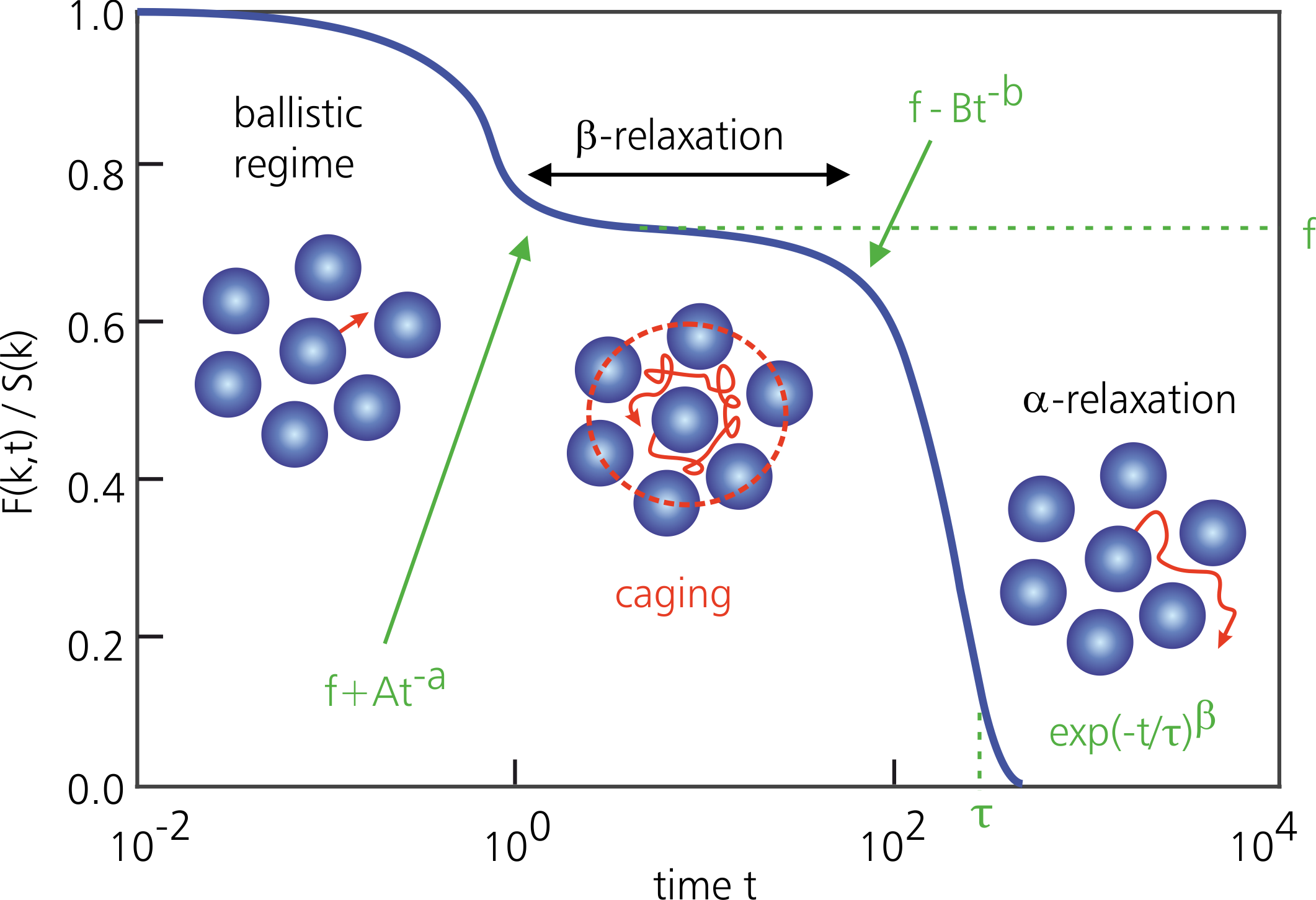}
  \end{center}
  \caption{
  \label{fig:Fkt_MCT}
  Typical MCT prediction for $F(k,t)$ of a supercooled liquid as a function of time,
  for a wavenumber $k=k_0$ that corresponds to the first peak of the static structure factor. 
  At very short times, particles undergo ballistic motion. At intermediate times, particles become
  transiently trapped in cages ($\beta$-relaxation) and $F(k,t)$ correspondingly
  remains approximately constant. Only at sufficiently long times will particles
  break free and full relaxation takes place ($\alpha$-relaxation).
  }
\end{figure}

A related success of MCT is its prediction of the \textit{cage effect} as a
microscopic mechanism for vitrification (Figs.\ \ref{fig:dynamics} and \ref{fig:Fkt_MCT}). 
Caging refers to the fact that, in a
supercooled liquid, particles become trapped in local cages formed by their
neighboring particles, preventing them from moving around as in a normal
liquid.  This is the molecular origin of the $\beta$-relaxation regime, which
is manifested as a plateau in $F(k,t)$.  As long as the material is on the
supercooled-liquid side of the transition, the particles will eventually manage
to escape their cages, but at and below the glass transition, the cage effect
keeps them trapped indefinitely. The only motion in the glassy state then
corresponds to a vibrational or rattling motion of the particles within their
confining cages. More mathematically, the cage effect emerges from MCT by
considering that the most prominent change in $S(k)$ upon supercooling occurs
at the main peak at wavenumber $k_0$, corresponding to length scales of
approximately one particle diameter. As a consequence, the first
intermediate scattering function that falls out of equilibrium at the glass
transition is $F(k_0,t)$, which in turn drives the freezing on all other
wavevectors. Notably, within MCT, the dominant structural length scale
governing vitrification thus remains on the order of only one particle
diameter, in stark contrast with conventional critical phenomena that are
usually accompanied by diverging, macroscopic length scales.
However, as will be described in Sec.\ \ref{sec:IMCT}, recent work suggests
that a diverging length scale also emerges within an extended ('inhomogeneous') version
of MCT that is related to the dynamic susceptibility $\chi_4(t)$.

Regardless of the molecular details of the material, which are contained in
$S(k)$, MCT also makes several general predictions for the relaxation dynamics
\cite{Gotze2008,Leutheusser1984,Kob2002,Reichman2005}.
Firstly, MCT predicts that close to the glass transition temperature $T_c$, the
relaxation time $F(k,t)$ will always diverge as a power law, $\tau \sim
(T-T_c)^{-\gamma}$.  Such a functional form is often in good agreement with
experiments and simulations in the mildly supercooled regime (using $\gamma$ as
a fit parameter), but generally breaks down closer to the experimental glass
transition (see e.g.\ Ref.\ \cite{Brambilla2009}. We will return to this point in the next subsection. Furthermore,
MCT predicts that the onset and decay of the $\beta$-relaxation regime, i.e.,
the plateau in $F(k,t)$ at intermediate times, are described by power laws of
the form $F(k,t) \sim f+At^{-a}$ and $f-Bt^b$, respectively, where $f$ is the
(constant) plateau height (Fig.\ \ref{fig:Fkt_MCT}).  Sufficiently close to $T_c$, the MCT exponents $a$
and $b$ are related as $\Gamma(1-a)^2/\Gamma(1-2a)=\Gamma(1+b)^2/\Gamma(1+2b)$,
where $\Gamma$ denotes the Gamma function.  This is an entirely non-trivial and
remarkable prediction that is fully consistent with experiments and
simulations. For the $\alpha$-relaxation regime, i.e., the final decay of
$F(k,t)$ on the liquid side of the transition, MCT predicts a stretched
exponential of the form $\exp(-t/\tau)^{\beta}$, with $0<\beta\leq 1$ (Fig.\ \ref{fig:Fkt_MCT}). This is
again in excellent agreement with experimental and simulation data, and
physically arises from the coupling of multiple density-mode relaxation
channels over different length scales, each relaxing on its own time scale.
Another success of MCT that has been verified experimentally is its prediction
of a time-temperature superposition principle, such that
$F(k,t)=\hat{F}(k,t/\tau(T))$, where $\hat{F}$ is a master function and
$\tau(T)$ is the $\alpha$-relaxation time.

Among the other celebrated results of MCT, we mention here its qualitative
prediction of complex reentrant effects in the behavior of sticky hard spheres
(particles with a hard repulsive core and short-ranged attractions) \cite{Pham2002} and
ultrasoft repulsive particles \cite{Berthier2010}, which exhibit glass-fluid-glass and
fluid-glass-fluid phases upon a monotonic increase in attraction stength and
density, respectively. In the case of sticky hard spheres, MCT has also
provided a qualitative explanation for the existence of the two distinct glass
phases in terms of different dominant length scales \cite{Pham2002}. Furthermore, the so-called
schematic version of MCT \cite{Bengtzelius1984,Leutheusser1984}, which is obtained by ignoring all wavevector
dependence in Eq.\ (\ref{eq:MCT}), is rigorously exact for certain classes of
spin-glass models with queched disorder (so-called $p$-spin spherical spin glasses), pointing toward a possible deep
connection between systems with quenched and self-generated disorder.  For a
more extensive overview of MCT results, we refer the reader to Refs.\
\cite{Kob2002} and especially \cite{Gotze2008}.

\subsection{Failures}
\begin{figure}
        \begin{center}
    \includegraphics[width=0.37\textwidth]{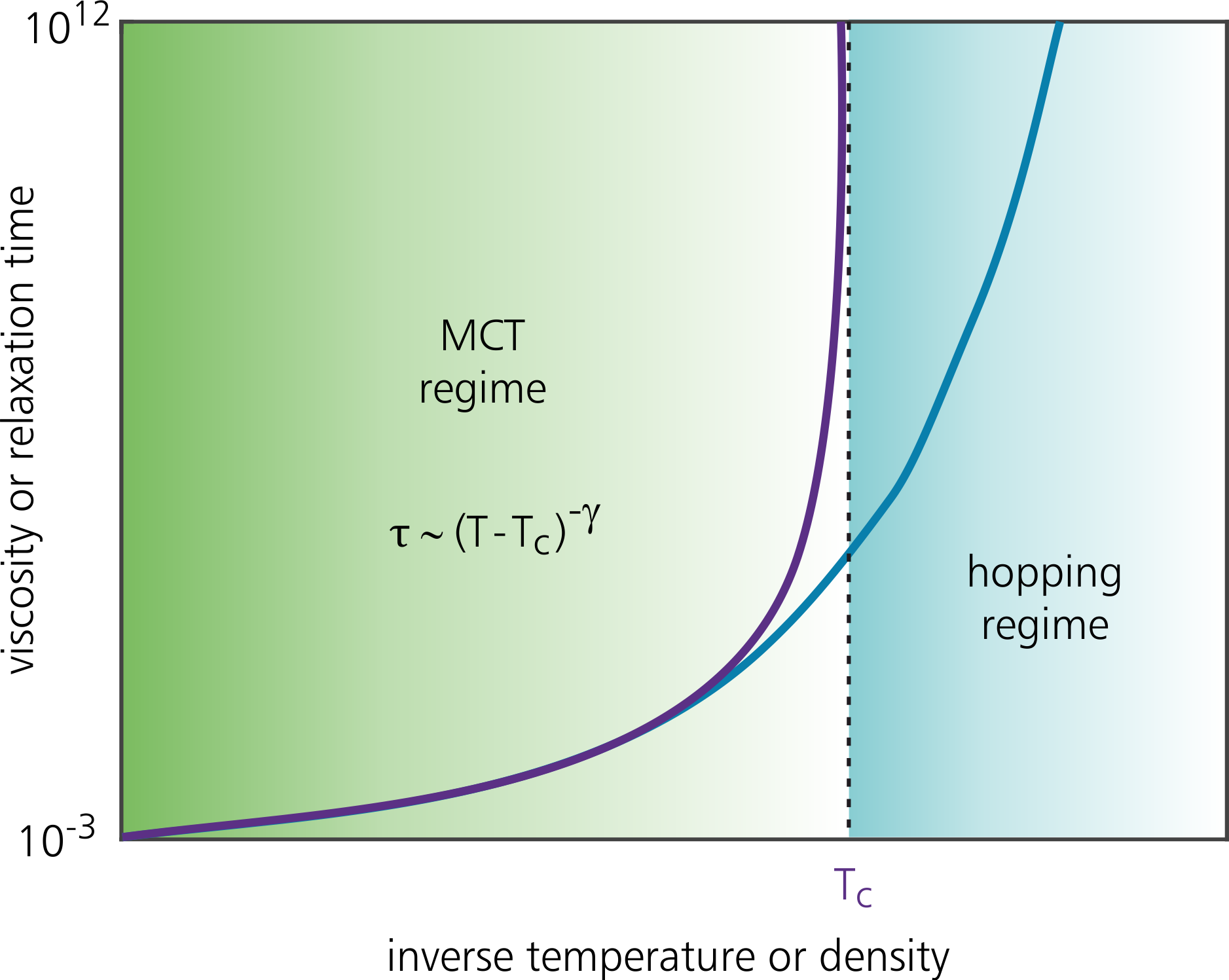}
  \end{center}
  \caption{
  \label{fig:MCThop}
  Typical MCT prediction (purple curve) and simulation result (blue curve) 
  for the dynamical slowdown of a glass-forming material as a 
  function of the control parameter. MCT generally predicts that
  the viscosity or relaxation time grows as a power law and diverges at the glass transition
  temperature $T_c$ or critical packing fraction $\phi_c$. In reality, a material tends to remain in the supercooled-liquid 
  phase at temperatures well below $T_c$ (or packing fractions above $\phi_c$), 
  which is attributed to so-called activated or hopping dynamics missing in standard MCT.
  The figure is based on Ref.\ \cite{Charbonneau2014}.
  }
\end{figure}

Even though MCT successfully predicts a glass transition, its most notable
failure is that the predicted glass transition temperature $T_c$ occurs at much
higher temperatures than the true experimental value $T_g$. Thus, the static
structure factor for which MCT predicts a glassy state corresponds in reality
to only a mildy supercooled liquid. In practice, the MCT predictions are often
rescaled such that $T_c$ coincides with $T_g$ \cite{Weysser2010}, but even with such a relative
comparison, MCT generally fails to accurate describe the dynamics in the deeply
supercooled regime. This discrepancy is attributed to MCT's
lack of ergodicity-restoring relaxation mechanisms that keep the experimental
system in the liquid phase well below $T_c$ (Fig.\ \ref{fig:MCThop}). Such mechanisms are generally
referred to as \textit{activated dynamics}, and are commonly identified with
particles 'hopping' out of their local cages to resist freezing \cite{Charbonneau2014}. MCT fails to
account for such hopping motion and thus strongly overestimates the degree of
caging--a feature that is believed to arise from its mean-field nature. In
practice, the predicted MCT transition at $T_c$ is therefore interpreted as a
crossover point where the dynamics changes into an activated form \cite{Biroli2009}. 
In Sec.\ \ref{sec:beyondMCT},
we will return to this point and address recent efforts to incorporate
activated dynamics directly into the theory.
We note that activated dynamics may also be incorporated via, e.g., the Random 
First Order Transition Theory (RFOT), which is a spin-glass-inspired framework 
that merges MCT with thermodynamics-based concepts \cite{Kirkpatrick1987,Kirkpatrick1987a}.
A description of RFOT falls outside the scope of the present work, but we refer the
interested reader to e.g.\ Refs.\ \cite{Biroli2009} and \cite{Kirkpatrick2014} for a
recent overview.

As mentioner earlier, MCT's prediction of a power-law divergence of the
relaxation time also breaks down in most experimental and simulated
glass-forming systems.  More generally, the fact that MCT always yields a power
law, regardless of the molecular composition of the material, also implies that
MCT has essentially no notion of the concept of fragility. At best, an MCT
power law may correctly describe the relaxation dynamics of \textit{fragile}
glass formers, but \textit{strong} glass formers exhibit a fundamentally
different, Arrhenius-type growth of the relaxation time. Indeed, an accurate
(first-principles) prediction of the fragility of a material on the sole basis
of its microscopic structure remains a major open challenge in the field
\cite{Tarjus2011}. Nonetheless, we note that MCT can predict other properties
of strong glass formers rather accurately, such as the wavevector-dependent
non-ergodicity parameter in the glassy phase \cite{Sciortino2001}.

MCT is also generally unable to account for the breakdown of the
Stokes-Einstein relation in the deeply supercooled regime.  This is again
attributed to the inherent mean-field character of the theory and the absence
of activated hopping dynamics \cite{Charbonneau2014}. Moreover, in its standard
formulation, MCT does not offer an explanation for the emergence of dynamic
heterogeneity, since MCT only predicts a single $F(k,t)$ for a given
wavevector, density, and temperature, and hence does not give access to
correlations in the \textit{fluctuations} of $F(k,t)$.  However, as discussed
in Sec.\ \ref{sec:IMCT}, an extension of the theory does allow for the calculation of a
quantity related to the dynamic susceptibility $\chi_4(t)$ and a corresponding growing (and
ultimately diverging) correlation length scale.  Furthermore, despite its
mean-field character, it was recently shown that MCT does \textit{not} become
exact in the mean-field limit of infinite dimensions for a system composed of
hard spheres \cite{Ikeda2010,Schmid2010,Maimbourg2016}, making it difficult to
rationalize the set of standard-MCT approximations in a simple physical manner.
Finally, since MCT is a purely dynamical theory, it cannot make any statements
about \textit{thermodynamic} properties such as the entropy. The latter is
believed to also play an important role in the process of glass formation, and
in particular may point toward an underlying thermodynamic transition that in
practice is masked by the dynamic transition. Nonetheless, it is possible that
MCT is implicitly aware of at least some changes in thermodynamic properties
through changes in the static structure factor \cite{Banerjee2014,Nandi2015}.

\section{Going beyond standard Mode-Coupling Theory}
\label{sec:beyondMCT}
Since standard MCT is not exact, as exemplified by the drawbacks and failures
discussed in the previous section, various attempts have been made in the last
few decades to improve the theory's predictive power for glassy dynamics.
Below we will summarize the most notable efforts to remedy at least some of
MCT's problems, including the formulation of 'Extended' MCT (EMCT) and
'Generalized' MCT (GMCT) to incorporate activated dynamics mechanisms, the
potential of GMCT to account for fragility and dynamic heterogeneity, and the
formulation of 'Inhomogeneous' MCT (IMCT) to predict dynamic susceptibilities.
Finally, we also briefly discuss recent generalizations of MCT to a new class
of soft condensed-matter systems referred to as \textit{active matter}. Such
active materials are composed of particles that can undergo autonomous motion
through the consumption of energy, and are now emerging as a new paradigm
to understand collective behavior seen in many living systems. The recent realization
that active particles can also vitrify into a glassy state has spurred the
formulation of various MCT frameworks for active matter, the development of which will be
reviewed in Sec.\ \ref{sec:active}.

\subsection{'Extended' Mode-Coupling Theory: incorporating couplings to currents}
The first attempts to remove the spurious MCT transition at $T_c$ were proposed
by Das and Mazenko in 1986 \cite{Das1986} and by G\"{o}tze and Sj\"{o}gren in 1987 \cite{Gotze1987}, only a few years
after the original formulation of standard MCT \cite{Bengtzelius1984,Leutheusser1984}.
Das and Mazenko employed a field-theoretic description, commonly referred to as
fluctuating nonlinear hydrodynamics, while G\"{o}tze and Sj\"{o}gren used a
projection-based formalism to improve the theory in the temperature regime near
and below $T_c$.  Both approaches amount to a perturbative treatment of
nonlinear couplings to certain current modes that are neglected in the standard
formulation of MCT, and which cut off the sharp MCT transition such that the
strict divergence of the relaxation time at $T_c$ is removed. This 'rounding
off' of the MCT transition was interpreted as a mechanism for activated or
hopping dynamics that would keep the material ergodic, i.e., in the supercooled
liquid phase, below $T_c$. The 2004 review by Das \cite{Das2004} provides an extensive
overview of this line of EMCT research.

However, more recent theoretical studies have argued on general physical
grounds that the invoked couplings to currents in EMCT cannot provide a
satisfactory explanation of activated dynamics, since these couplings should
always become negligible close to a glass transition \cite{Cates2006}.  Moreover,
Andreanov \textit{et al.}\ \cite{Andreanov2006} suggested that the fluctuating nonlinear
hydrodynamics approach employs an incorrect treatment of time-reversal
symmetry. Another argument that casts doubt on the general applicability of
EMCT is the fact that experimental and numerical simulation studies have
unambiguously established that materials obeying Newtonian and Brownian
(stochastic) dynamics exhibit the \textit{same} deviations from standard-MCT
behavior, despite their differences in microscopic dynamical details.  This
suggests that the physical mechanisms governing activated behavior below $T_c$
have a universal origin in both molecular (Newtonian) fluids and colloidal
(Brownian) systems.  Since the current modes introduced in
EMCT \textit{cannot be properly defined} in Brownian systems \cite{Szamel2013}, the proposed EMCT
mechanism may thus only apply to materials undergoing Newtonian dynamics.
Hence, it appears likely that EMCT cannot offer a rigorous, universal remedy
for the lack of ergodicity-restoring activated dynamics within the standard MCT framework.

\subsection{'Generalized' Mode-Coupling Theory: towards an exact equation for the memory function}
\begin{figure}
        \begin{center}
    \includegraphics[width=0.49\textwidth]{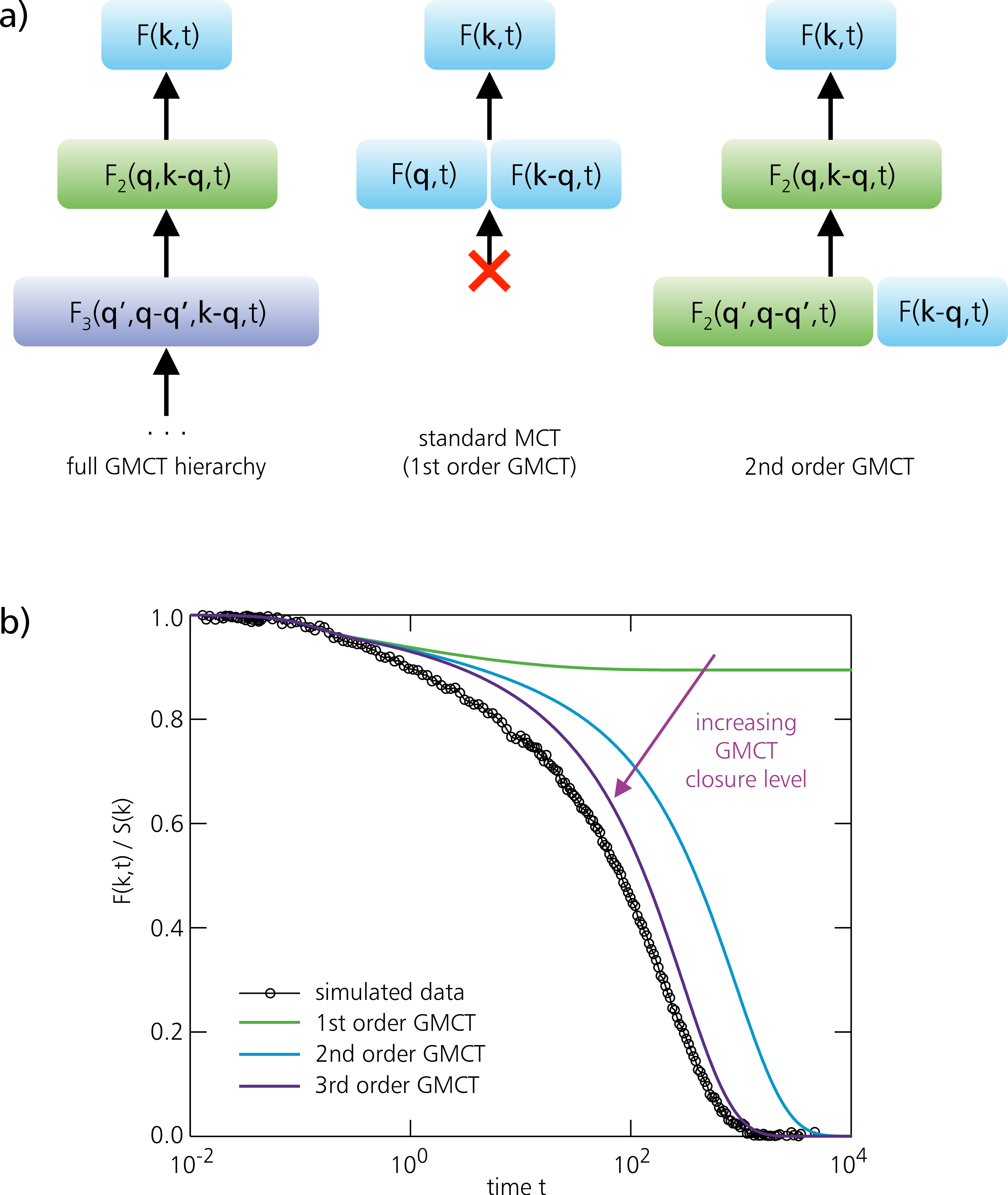}
  \end{center}
  \caption{
  \label{fig:GMCT}
  (a) Graphical illustration of the GMCT hierarchy. GMCT seeks to systematically avoid 
  the uncontrolled MCT factorization approximation by developing a new, and formally exact, 
  equation of motion for the unknown memory function. This equation in turn is governed 
  by a new memory function, which is controlled by another memory function, etc.
  Standard MCT corresponds to the lowest-order (self-consistent) closure of this hierarchy.
  (b) Microscopic GMCT predictions for $F(k,t)$ \cite{Janssen2015a} compared to numerical data obtained from 
  computer simulations \cite{Weysser2010} for a system of quasi-hard spheres at
  packing fraction $\phi=0.570$ and wavenumber $kd=7.4$.  These results
  indicate that the GMCT hierarchy apparently converges and that the theory
  becomes more quantitatively accurate as the closure level is increased.
  The figure is adapted from Ref.\ \cite{Janssen2015a} with permission.
  }
\end{figure}

An alternative route to rigorously improve MCT was put forward by Szamel in
2003 \cite{Szamel2003}.  This approach, referred to as Generalized MCT or GMCT, seeks to
systematically avoid the second main approximation of standard MCT, i.e., the
uncontrolled factorization of the four-point density correlations appearing in
the memory function. To this end, a new and formally exact equation of
motion is developed \textit{for the four-point correlation functions themselves} (again
by applying the Mori-Zwanzig projection formalism of Sec.\ \ref{sec:MZ}, this time using
the basis of pair densities $\rho(\mathbf{k}_1,t)
\rho(\mathbf{k}_2,t)$ as the 'relevant' variables).  The new equation is
governed by another memory function that, to leading order, is controlled by
six-point density correlation functions, which in turn are dominated
by eight-point correlations, etc. Hence, by repeatedly developing a new
equation of motion for the new memory function, a \textit{hierarchy} of coupled
equations emerges, in which the uncontrolled factorization approximation may be
applied at an arbitrary level to close the set of equations. This GMCT scheme
thus allows, in principle, for a \textit{systematic} delay of the closure
approximation and, notably, remains based entirely on first principles (see Fig.\ \ref{fig:GMCT}).  

Szamel \cite{Szamel2003} and Wu and Cao \cite{Wu2005} showed that GMCT hierarchies factorized at the level of
six- and eight-point correlation functions, respectively, indeed bring the
predicted glass transition density systematically closer to the empirical value
for a system of colloidal hard spheres. More recent work \cite{Janssen2015a} also established that
the full time-dependent microscopic dynamics for a quasi-hard-sphere glass
former is systematically improved by GMCT. In fact, fit-parameter-free
third-order GMCT calculations could achieve full quantitative agreement for
$F(k,t)$ up to the moderately supercooled regime, at densities where standard
MCT would already predict a spurious glass transition \cite{Janssen2015a}. 
 Furthermore, within a
simplified schematic (wavevector-independent) GMCT model, Mayer \textit{et
al.}\ \cite{Mayer2006} showed analytically that the sharp MCT glass transition can be completely
removed when avoiding the closure approximation altogether, i.e., when applying
infinite-order GMCT. Even though all GMCT studies to date still rely on several
approximations--including the neglect of 'projected' dynamics in the memory
functions (Sec.\ \ref{sec:MCT}) and the factorization of all \textit{static} correlation
functions into products of $S(k)$'s--, the good agreement so far with 
computer simulations and experiments, as well as the apparent convergent
behavior of the hierarchy \cite{Janssen2015}, suggest that GMCT offers a promising
first-principles path towards systematic MCT improvement.  In particular, it
appears that higher-order GMCT captures at least some aspects of activated
dynamics to keep the material ergodic at temperatures below $T_c$, consistent
with empirical observations. Importantly, we note that GMCT is applicable to
both Newtonian and Brownian systems, and therefore also holds the potential to
offer a more universal picture of glassy dynamics.

In addition to accounting for some kind of ergodicity-restoring processes below
$T_c$, GMCT might also provide a suitable framework to describe fragility. The
work of Mayer \textit{et al.}\ \cite{Mayer2006} revealed that, within their particular schematic model,
infinite-order GMCT predicts an \textit{exponential} growth of the relaxation
time, fundamentally distinct from the fragile power-law behavior of standard
MCT.  In later studies, we demonstrated that other schematic GMCT models may
also give rise to other functional forms of relaxation-time growth, ranging
from fragile super-Arrhenius to strong (sub-)Arrhenius behavior, depending on
the choice of schematic parameters \cite{Janssen2014}. Although these simplified GMCT models
inherently lack any wavevector dependence, and therefore cannot make detailed
predictions for any structural glass former with a realistic $S(k)$, they
suggest that higher-order GMCT has at least the \textit{mathematical
flexibility} to account for different fragilities. This is notably different
from standard MCT, which is mathematically only capable of predicting power-law
growth close to the transition.  It remains to be tested whether the fully
microscopic (wavevector-dependent) version of GMCT will indeed be able to
account for different degrees of fragility, given solely the static structure
factors $S(k)$ (and possibly higher-order static correlation functions) of
strong and fragile materials as input. It might be tempting to assume that,
with increasing closure level, the GMCT predictions should become more
accurate, but let us reiterate that the current formulation of GMCT still
relies on several approximations, and it is still unclear how the remaining
assumptions ultimately affect the dynamics.

Finally, we note that \textit{by construction}, higher-order GMCT also makes
microscopic predictions for the (approximate) dynamics of unfactorized four-point density
correlations \cite{Janssen2015a}.  Although these functions are not exactly equivalent to the dynamic
susceptibility $\chi_4(t)$, they should nonetheless be able to provide insight
into dynamic heterogeneities, since they essentially describe particle
correlations over two points in time and at least two points in space. Hence,
GMCT may also offer a suitable starting point to study dynamic heterogeneity,
as well as the breakdown of the Stokes-Einstein relation in supercooled
liquids, from a strictly first-principles perspective. We expect this avenue of
research to be explored in the coming years.

\subsection{'Inhomogeneous' Mode-Coupling Theory: a measure for dynamic heterogeneity}
\label{sec:IMCT}
As noted earlier, standard MCT seeks to describe the 'average' $F(k,t)$ for a
given set of wavevectors and system parameters, but does not give immediate
access to the \textit{fluctuations} of $F(k,t)$ that are encoded in the dynamic
susceptibility $\chi_4(t)$ \cite{Biroli2004}. Hence, standard MCT cannot make
direct predictions about dynamically heterogeneous behavior, which is generally
revealed as a growing peak in $\chi_4(t)$. There is, however, an indirect way
to extract a dynamic susceptibility from MCT by incorporating an
\textit{external field} into the theory--a framework referred to as
Inhomogeneous MCT or IMCT. The idea of IMCT is to measure the dynamic response
of the intermediate scattering function $F(k,t)$ to changes in the external
field; this response amounts to a \textit{three-point} dynamic density
correlation function $\chi_3(t)$. The IMCT study of Biroli \textit{et al.}\
\cite{Biroli2006} argues that the \textit{induced} fluctuations by the external
field are intimately related to the \textit{spontaneous} fluctuations described
by $\chi_4(t)$, and hence the susceptibility $\chi_3(t)$ should behave in a
similar manner as the four-point function $\chi_4(t)$. 

Biroli \textit{et al.}\ found that $\chi_3(t)$ grows upon
approaching the dynamical MCT transition, and in fact diverges at the critical
temperature $T_c$. Furthermore, a correlation length $\xi$ could be defined--a
measure, perhaps, for the size of cooperatively rearranging particles in the
supercooled regime--, that grows as $\xi \sim |T-T_c|^{-\nu}$ with a critical
exponent of $\nu=1/4$. Notably, IMCT also predicts that this
length scale governs both the $\alpha$- and $\beta$-relaxation regimes. This
suggests that the traditional picture of caging in the $\beta$-regime, commonly
interpreted as the rattling of particles in local cages formed by their nearest
neighbors (see \ref{sec:successes}), is actually more subtle; rather, IMCT
implies that these cages become more and more collective as the MCT transition
is approached. However, it should be noted that the predictions of IMCT are not
generally in quantitative agreement with empirical results. For example,
numerical simulations for a model glass former composed of Lennard-Jones
particles indicate a growth of $\xi$ (extracted from the numerical $\chi_4(t)$)
with a fitted exponent of $\nu\approx0.5$, and suggest that the length scale
predicted by IMCT does not necessarily describe the true size of the
correlated spatial domains relevant in real glass-forming materials
\cite{Karmakar2009}.  On the other hand, simulations for another model glass
former (the so-called Gaussian core model, which is believed to behave more as
a mean-field system) have revealed that the predicted IMCT scaling of
$\chi_3(t)$ is in good quantitative agreement with the numerical $\chi_4(t)$
\cite{Coslovich2016}, implying that IMCT constitutes at least in some sense a
suitable mean-field framework for glassy dynamics. The question to what extent,
and under which conditions, IMCT can offer an accurate description of dynamic
heterogeneity, and how the IMCT predictions relate to, e.g., the four-point
dynamic correlations emerging from GMCT, still remains to be established.

\subsection{Mode-Coupling Theories for active matter}
\label{sec:active}
We end this review with a very recent development in the field, namely the
study of \textit{active matter}. Active materials consist of particles that can
convert energy into autonomous motion, rendering them out of
thermodynamic equilibrium \textit{at the single-particle level} \cite{Marchetti2013}. Such particle
activity can lead to rich self-organizing behavior, as exemplified in nature
by, e.g., the collective motion of living cells and the flocking of birds.
During the last decade, numerous synthetic active systems have also become
available \cite{Bechinger2016}, spurring the development of theoretical approaches to describe the
emergent behavior in these non-equilibrium materials. In particular, it was
found that dense active matter can also exhibit properties of supercooled
liquids and vitrifying colloidal suspensions \cite{Angelini2011,Henkes2011,Ni2013,Sadati2014,Berthier2014,Pilkiewicz2014,
Szamel2015a,Bi2015,Delarue2016,Bi2016,Yazdi2016,Berthier2017,Janssen2017}, including slow structural
relaxation, dynamic heterogeneity, varying degrees of fragility, and the
ultimate formation of a kinetically arrested, amorphous solid state.

Here we briefly discuss recent extensions of standard MCT to describe the
glassy dynamics in active materials. Since many synthetic active particles
are composed of colloids undergoing active Brownian motion, all active
versions of MCT to date are based on the Smoluchowski formalism for Brownian
systems, rather than the Newtonian description for molecular fluids discussed in
Sec.\ \ref{sec:deriv}. We note, however, that \textit{continuum} descriptions 
of active matter, such as those for active liquid crystals, are usually derived from 
Newtonian-based fluid mechanics \cite{Marchetti2013}.

The first MCT approach to active glasses was presented by Farage and Brader in
2014 \cite{Farage2014}. In this work, they considered so-called active Brownian
particles (ABPs) that move with a constant self-propulsion speed in a random
direction, subject to translational and rotational Brownian motion.  The
authors assumed that a \textit{single}, non-interacting ABP behaves
\textit{effectively} as a passive colloid, but with a higher effective
diffusion constant. This approximation was subsequently used to derive an
effective Smoluchowski operator for the collective dynamics of a dense ensemble
of active particles. In essence, this effective-diffusion approach amounts to
the removal of explicit rotational degrees of freedom. 
The resulting MCT approach yields a modified version
of Eq.\ (\ref{eq:MCT}), in which both the frequency term and the memory
function acquire an activity-dependent prefactor. The main outcome of this MCT
study is that the addition of particle activity can soften (i.e., decrease the
non-ergodicity parameter) and eventually melt a passive glass, and shift the
glass transition toward higher densities, These findings are also in qualitative agreement
with computer simulations of a similar active material composed of
self-propelling Brownian hard particles \cite{Ni2013,Berthier2014}. The MCT approach of 
Farage and Brader was later also extended by Ding \textit{et al.}\ \cite{Ding2015} to mixtures of active and passive particles.
 
A different and more extensive active-matter study was performed by Szamel and
co-workers \cite{Szamel2015a,Szamel2016}. Here, the authors modeled active particles by an Ornstein-Uhlenbeck
stochastic process, characterized by an effective temperature that quantifies
the strength of the active forces, and a persistence time that describes the
duration of persistent self-propelled motion. In this model, particle motion is thus
described as a persistent random walk.
Within their framework, the
self-propulsion is first integrated out before applying the projection-operator
method and MCT-like approximation; this approach essentially assumes that particle positions
evolve on a time scale much larger than the time scale needed for reorientation of the activity direction,
somewhat akin to the effective-diffusion assumption of Farage and Brader \cite{Farage2014}. 
 An important difference between the active MCT of Szamel \textit{et al.}\ 
and previous MCT studies is that not only the static structure
factor--i.e., static correlations between particle positions--should be given
as input to the theory, but also static correlations between particle
\textit{velocities}.  Contrary to the behavior of ABPs, it was found that the
incorporation of activity can both enhance and suppress glass formation: for
small persistence times, the active fluid relaxes faster than a passive system
at the same effective temperature, but for large persistent times the active
material becomes \textit{more glass-like} compared to the passive reference
system.  This non-monotonic dependence of the relaxation time was observed both
in the MCT analysis and in computer simulations, and was attributed to the
competition between increasing velocity correlations (which speed up the
dynamics) and increasing structural correlations (which slow down the
dynamics) \cite{Szamel2015a}.  For sufficiently large persistence times, it was found that the
fitted MCT glass transition temperature increases monotonically with increasing
persistence time, suggesting that--at least within this active-matter
model--vitrification occurs more easily as the material becomes more active.
An MCT-based scaling analysis for this type of active-matter system was later 
performed by Nandi and Gov \cite{Nandi2017a}.

Feng and Hou \cite{Feng2017} subsequently studied a quasi-equilibrium thermal version of the active
Ornstein-Uhlenbeck model of Szamel and co-workers, which additionally accounts
for thermal translational noise.  Their MCT derivation differs from the
approach taken by Szamel \cite{Szamel2016}, however: it is valid only for sufficiently small
persistence times (since it relies on a perturbative expansion), and does not
require explicit velocity correlation functions to be given as input.  Rather,
their active-MCT dynamics is governed by an averaged diffusion constant
$\bar{D}$ and a non-trivial steady-state structure function $S_2(k)$, which
both depend on the effective temperature and density of the system, as well as
on the persistence time of the active particles. The coefficient $\bar{D}$ and
$S_2(k)$ should both be given as additional input to the theory in order to
predict $F(k,t)$. It was found that the critical density at which the glass
transition takes place shifts to larger values with increasing magnitude of the
self-propulsion force or effective temperature, and that the critical effective
glass temperature increases with the persistence time. In the limit of a
vanishing persistence time, the theory naturally yields the expected result for
a simple passive Brownian system \cite{Feng2017}.

Very recently, Liluashvili, \'{O}nody, and Voigtmann \cite{Liluashvili2017} formulated the first MCT
for ABPs in which both the translational and rotational degrees of freedom are
treated on an equal footing. That is, rather than seeking to reduce the active
material to a near-equilibrium system, the rotational degrees of freedom
governing the reorientation of the active forces are now explicitly coupled to
the translational motion.  This approach thus avoids the effective-diffusion
assumption (which in principle may be valid only at low densities and sufficiently long
times), and the resulting dynamics now also depends non-trivially on the
rotational diffusion constant.
The only required material-dependent input for this active MCT is the
passive-equilibrium static structure factor. An important outcome of this study
is the three-dimensional fluid-glass phase diagram for hard ABPs as a function
of packing fraction, self-propulsion speed, and rotational diffusion constant.
It was shown that this surface cannot be collapsed onto a single line in the
two-dimensional plane, highlighting the importance of treating the rotational
degrees of freedom explicitly. Indeed, depending on the density of the active
material, separate regimes could be identified that are dominated either by
translational or reorientational motion.  As in the study of Farage and Brader \cite{Farage2014},
and in agreement with computer simulations \cite{Ni2013,Berthier2014}, it was also found that activity
generally makes hard-sphere systems more fluid-like and consequently shifts the
glass transition to higher packing fractions.  Notably, this active
fluidization effect grows \textit{monotonously} with increasing persistence
time or inverse rotational diffusion constant, in contrast with the findings of
Szamel and co-workers \cite{Szamel2015a}. This difference is attributed to the absence of thermal
Brownian noise in the model of Szamel \textit{et al.}: in the limit of
infinitely large persistence (vanishing rotational diffusion), active particles
can block themselves and produce a glassy state, while the finite thermal
diffusive motion in ABPs will make such blocking ineffective \cite{Liluashvili2017}.

\section{Conclusions and outlook}
This review has sought to provide a brief overview of the main phenomenology of
glassy dynamics, and of its theoretical description using Mode-Coupling
Theory--currently the only theory of the glass transition that is based
entirely on first principles. We have focused mainly on the behavior of the
density correlation function $F(k,t)$ as a probe of the microscopic dynamics
associated with vitrification. In the normal liquid phase, this correlation
function rapidly decays to zero, but at the glass transition it fails to decay
on any practical time scale, marking the onset of rigidity and providing an
order parameter for the transition. Upon approaching the glass transition
temperature, several complex features become visible in the dynamics, such as a
transient plateau and stretched exponential behavior in $F(k,t)$, a breakdown
of the Stokes-Einstein relation, and the emergence of dynamical
heterogeneity--the latter being associated with increasingly large fluctuations
in $F(k,t)$.  Remarkably, during the process of glass formation, the
microscopic \textit{structure} of the material, as probed by e.g.\ the radial
distribution function $g(r)$ or static structure factor $S(k)$, undergoes only
very minor changes, yet the viscosity and dynamic relaxation time increase by
many orders of magnitude. It is this seemingly paradoxical discrepancy between
structure and dynamics that makes the glass transition a notoriously difficult
problem in theoretical physics.

MCT offers a first-principles-based framework to account for at least some
aspects of glassy dynamics.  Its starting point is the \textit{exact} equation
of motion for $F(k,t)$; through a series of (partly uncontrolled)
approximations, MCT subsequently provides a self-consistent equation for
$F(k,t)$ that can be solved numerically using only the static structure factor
as input. As such, the theory makes a set of detailed predictions for the full
microscopic relaxation dynamics of a glass-forming material as a function of
time, wavevector, temperature, and density, on the sole basis of simple
structural information.  Among its notable successes is the qualitative
prediction of a glass transition, a physically intuitive picture for glass
formation in terms of the cage effect, and the correct prediction of several
highly non-trivial scaling behaviors in $F(k,t)$. However, MCT is generally not
\textit{quantitatively} accurate, and cannot account properly for the concept
of fragility, the violation of the Stokes-Einstein relation, and the emergence
of dynamic heterogeneity.

The shortcomings of MCT might be remedied using (first-principles-based)
extensions of the theory, such as Generalized MCT and Inhomogeneous MCT. The
first studies in this direction show that GMCT can indeed offer a more
quantitative description of the $F(k,t)$ dynamics and can potentially describe
fragility, while IMCT offers a framework to qualitatively account for dynamic
heterogeneity.  However, GMCT still relies on several approximations such as
the neglect of certain wavevector-dependent density correlations, and IMCT
provides--just like standard MCT--only a mean-field description of glassy
dynamics. Hence, more work will be needed to establish how successful these
theoretical approaches are in ultimately achieving a fully correct
first-principles description of glassy dynamics. 
 
A more recent addition to the palette of Mode-Coupling theories involves the
study of non-equilibrium active matter. In the last few years, several MCT
frameworks have been developed to describe glassy dynamics in active materials
that are composed of self-propelled particles. Not only can these theories
offer new insight into the behavior of dense assemblies of synthetic active
colloids, but they might also shed new light on glassy and jamming phenomena in
living cell tissues. Similar to how standard MCT has shaped our understanding
of passive glass-forming materials over the last few decades, it can be
expected that active MCT will also contribute to our understanding of
disordered active and living materials from a statistical-physics-based and
purely first-principles perspective.

In conclusion, despite the fact that Mode-Coupling Theory is not exact, it does
provide a suitable--and in some cases remarkably accurate--foundation for the
study of glassy dynamics in amorphous materials.  The theory also offers ample
opportunity for new research aimed towards a complete and ultimately rigorously
exact description of the glass transition, as well as for the study of emergent
new classes of materials such as active matter.  We expect future work to be
directed toward these exciting avenues of research.

\acknowledgments
It is a pleasure to thank David Reichman, Grzegorz Szamel, J\"{u}rgen Horbach, Thomas Voigtmann, Hartmut L\"{o}wen,
Matthias Fuchs, J\"{o}rg Baschnagel, Jean Farago, Atsushi Ikeda, and 
Peter Mayer for many interesting and enlightening discussions.

%

\end{document}